\documentclass[reprint,aps,prd,superscriptaddress]{revtex4-2}

\usepackage{bm}
\usepackage{amsmath}
\usepackage{amsfonts}
\usepackage{graphicx}

\begin{document}

\title{Acceleration of Ultrahigh Energy Particles from Fast Radio Bursts}

\affiliation{State Key Laboratory of Dark Matter Physics,Key Laboratory for Laser Plasmas (MoE), School of Physics and Astronomy, Shanghai Jiao Tong University, Shanghai 200240, China}
\affiliation{Tsung-Dao Lee Institute, Shanghai Jiao Tong University, Shanghai 201210, China}
\affiliation{Collaborative Innovation Center of IFSA, Shanghai Jiao Tong University, Shanghai 200240, China}
\author{Lin Yu}
\affiliation{State Key Laboratory of Dark Matter Physics,Key Laboratory for Laser Plasmas (MoE), School of Physics and Astronomy, Shanghai Jiao Tong University, Shanghai 200240, China}
\affiliation{Collaborative Innovation Center of IFSA, Shanghai Jiao Tong University, Shanghai 200240, China}
\author{Tianxing Hu}
\affiliation{State Key Laboratory of Dark Matter Physics,Key Laboratory for Laser Plasmas (MoE), School of Physics and Astronomy, Shanghai Jiao Tong University, Shanghai 200240, China}
\affiliation{Collaborative Innovation Center of IFSA, Shanghai Jiao Tong University, Shanghai 200240, China}
\author{Zhiyu Lei}
\affiliation{State Key Laboratory of Dark Matter Physics,Key Laboratory for Laser Plasmas (MoE), School of Physics and Astronomy, Shanghai Jiao Tong University, Shanghai 200240, China}
\affiliation{Collaborative Innovation Center of IFSA, Shanghai Jiao Tong University, Shanghai 200240, China}
\author{Xiangyan An}
\affiliation{Tsung-Dao Lee Institute, Shanghai Jiao Tong University, Shanghai 201210, China}
\affiliation{State Key Laboratory of Dark Matter Physics,Key Laboratory for Laser Plasmas (MoE), School of Physics and Astronomy, Shanghai Jiao Tong University, Shanghai 200240, China}
\affiliation{Collaborative Innovation Center of IFSA, Shanghai Jiao Tong University, Shanghai 200240, China}
\author{Dong Wu}
\affiliation{State Key Laboratory of Dark Matter Physics,Key Laboratory for Laser Plasmas (MoE), School of Physics and Astronomy, Shanghai Jiao Tong University, Shanghai 200240, China}
\affiliation{Collaborative Innovation Center of IFSA, Shanghai Jiao Tong University, Shanghai 200240, China} 
\author{Suming Weng}
\affiliation{State Key Laboratory of Dark Matter Physics,Key Laboratory for Laser Plasmas (MoE), School of Physics and Astronomy, Shanghai Jiao Tong University, Shanghai 200240, China}
\affiliation{Collaborative Innovation Center of IFSA, Shanghai Jiao Tong University, Shanghai 200240, China}
\author{Min Chen}
\affiliation{State Key Laboratory of Dark Matter Physics,Key Laboratory for Laser Plasmas (MoE), School of Physics and Astronomy, Shanghai Jiao Tong University, Shanghai 200240, China}
\affiliation{Collaborative Innovation Center of IFSA, Shanghai Jiao Tong University, Shanghai 200240, China}
\author{Jie Zhang}
\affiliation{State Key Laboratory of Dark Matter Physics,Key Laboratory for Laser Plasmas (MoE), School of Physics and Astronomy, Shanghai Jiao Tong University, Shanghai 200240, China}
\affiliation{Tsung-Dao Lee Institute, Shanghai Jiao Tong University, Shanghai 201210, China}
\affiliation{Collaborative Innovation Center of IFSA, Shanghai Jiao Tong University, Shanghai 200240, China}
\author{Zhengming Sheng}
\email{zmsheng@sjtu.edu.cn}
\affiliation{State Key Laboratory of Dark Matter Physics,Key Laboratory for Laser Plasmas (MoE), School of Physics and Astronomy, Shanghai Jiao Tong University, Shanghai 200240, China}
\affiliation{Tsung-Dao Lee Institute, Shanghai Jiao Tong University, Shanghai 201210, China}
\affiliation{Collaborative Innovation Center of IFSA, Shanghai Jiao Tong University, Shanghai 200240, China}

\date{\today}

\begin{abstract}
Two extreme events in the universe, fast radio bursts (FRBs) and cosmic rays (CRs), could be correlated, where FRBs with extreme field strength near their sources may contribute to CRs. This study investigates localized particle acceleration driven by FRB-like ultra-relativistic electromagnetic pulses {in an electron--positron--ion plasma system}. It is found ultra-high energy neutral plasma sheets form constantly via the front erosion of an FRB pulse. There are two regimes of ion acceleration depending upon the field strength and the plasma density: the piston regime driven by the Lorentz force of the pulse, and the wakefield regime dominated by charge separation field. The predicted energy scalings align well with particle-in-cell simulations. A power-law energy spectrum with an index close to the CRs naturally emerges during FRBs expansion outward. Detecting high-energy particles possibly produced by FRBs enables deeper insights into their origins and promotes the development of multi-messenger astronomy.
\end{abstract}

\maketitle

\section{Introduction}
Fast radio bursts (FRBs) are the most extreme coherent electromagnetic radiation observed to date from the universe \cite{RN1,RN2,RN4,RN5}. Observational data are rapidly accumulating \cite{RN64,RN88,RN43,RN60,RN61,RN62}, yet the physical origins of these bursts remain unclear. Extensive theoretical studies have explored their emission mechanisms and how propagation effects shape the observed signals, including their temporal, spectral, and polarization properties \cite{RN118,RN59,RN3,RN65,RN66,RN67}. FRBs have also been leveraged as powerful cosmological probes \cite{RN58,RN68,RN69,RN70}, with an estimated all-sky rate exceeding $10^4$ per day \cite{RN5}. Their multi-wavelength and multi-messenger counterparts, including X-rays \cite{RN72, RN89}, gamma-rays \cite{RN73}, neutrinos \cite{RN74}, and gravitational waves \cite{RN75, RN96}, have been observed, providing crucial insights into their origins \cite{RN77,RN78,RN79}. To the best of our knowledge, there has been no discussion on the possible connection between FRBs and cosmic rays (CRs).

CRs represent another extreme cosmic phenomena. In particular, the origin of ultra-high-energy cosmic rays (UHECRs) with energies extending beyond 100 EeV remains a mystery. Diffusive shock acceleration (DSA) is the most common shock acceleration to explain UHECRs, notable for naturally producing a power-law energy spectrum similar with observations \cite{RN9,RN10}. However, relativistic DSA is incapable of accelerating proton to 100 EeV \cite{RN11}. Non-relativistic DSA can potentially achieve ultra-high energy particle acceleration, but only under special shock modes and plasma conditions \cite{RN11,RN12}. Research on UHECRs remains challenging due to the difficulty in detecting the proposed sources. Multi-wavelength and multi-messenger observations, including very-high energy (VHE) neutrinos and ultra-high energy (UHE) gamma rays, have provided deeper insights into the underlying mechanisms, but these messengers are also constrained by the very low statistics of detectable events \cite{RN13,RN14,RN15}.

\begin{figure}[t] 
	\centering 
	\includegraphics[scale=0.9]{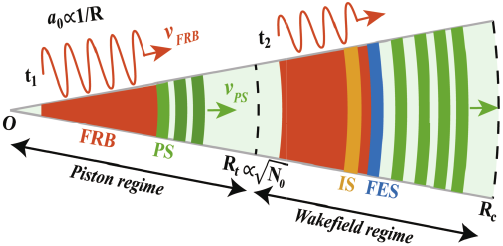}
	\caption{{Two regimes of particle acceleration driven by an expanding ultra-relativistic FRB pulse.} Ion is accelerated in the piston regime within a radius $R_t$, where energetic plasma sheets (PS) composed of electrons and ions are accelerated directly by the FRB pulse mediated by the pulse front erosion. Ion acceleration switches to the wakefield regime for a weakened pulse within the range $(R_t, R_c)$, where the front electron sheets (FES) are formed first at the eroded front of the pulse and subsequently the high electrostatic fields build up and accelerate the ion sheets (IS) until the two kinds of sheets coincide to form quasi-neutral PS. The PS move faster than the FRB ($v_{PS} \sim c > v_{FRB}$) and continuously form at the FRB front during its propagation.
	\label{fig:fig0}} 
\end{figure}

Thanks to their extremely high luminosities, FRBs are unique high-field coherent electromagnetic waves near their sources \cite{RN5,RN82,RN26}, far surpassing the intensity of the lasers currently available in laboratories. Over the past three decades, extensive research in laser-plasma laboratories has investigated particle acceleration driven by ultra-intense laser pulses. Charged particles can be efficiently accelerated either directly by a laser pulse with specific configurations \cite{RN16,RN17,RN18,RN19,RN20} or indirectly through high-amplitude electron plasma wakefields excited by the driver pulses \cite{RN21,RN22,RN51}. In astrophysical contexts, the electromagnetic radiation from pulsars can directly accelerate single-particle to ultra-high energy \cite{RN81}. It has also been proposed that the relativistic Alfvenic wave pulse propagating in the jet around a black hole could drive wakefield acceleration \cite{RN23,RN50,RN24}.

In this work, we theoretically demonstrate the remarkable acceleration capabilities of FRBs near their sources, suggesting their potential as cosmic-ray accelerators, even capable of producing UHECRs. In the ultra-relativistic regime, where the normalized field strength  $a_0 = {eE_0}/{m_ec\omega} > 1000$, two distinct regimes of particle acceleration are identified as an FRB pulse expands in plasma, as shown schematically in Fig.$\,$\ref{fig:fig0}. {Ions are accelerated directly by the Lorentz force of the pulse within certain distance $R_t$ (the piston regime). Beyond $R_t$, the acceleration enters the wakefield regime.} These regimes exhibit different energy scalings for accelerated ions.

\section{FRBs as ultra-relativistic electromagnetic waves near their sources}
The observed FRBs exhibit isotropic electromagnetic energies $W_\mathrm{iso} \approx 10^{35} \sim 10^{43}$ erg, durations $T \approx 1$ ms, and frequencies $f = \omega/2\pi$ ranging from 0.1 to 10 GHz \cite{RN5}. Near its source, an FRB pulse exhibits an extremely high electric field strength, given by $E_0 = \sqrt{W_\mathrm{iso} / cTR^2}$, {where $R$ denotes the radial distance from the geometric origin of the FRB pulse}, $c$ is the light speed.  At the frequency of $f = 1$ GHz, the normalized $E_0$ can be calculated as
\begin{equation}
	a_0 = \frac{eE_0}{m_ec\omega} \approx  {5.1\times{10}^8} %
	\left( \frac{W_\mathrm{iso}}{10^{40}\,\mathrm{erg}} \right) ^{1/2} %
    {\left( \frac{R}{1\,\mathrm{km}} \right) ^{-1}},  \label{equ:equ1}
\end{equation}
with $e$ and $m_e$ the unit charge and the rest mass of electrons, respectively. {The parameter $a_0$ decreases as the pulse expands outward, where its upper limit is governed by the underlying radiation mechanisms \cite{RN85,RN86}. In coherent bunch radiation models \cite{RN5,RN85,RN87}, electron bunches confined to a volume $V \sim cT\Gamma^2\lambda_0^2$ emit an FRB pulse into a solid angle $\Omega \sim \pi/\Gamma^2$, where $\lambda_0 = c/f$ is the FRB wavelength and $\Gamma \approx$ 100 -- 1000 denotes the Lorentz factor of the bunches. The local energy density is $\mathcal{W}_\mathrm{FRB} \sim W_\mathrm{iso}\Omega/4\pi V$. This yields an upper limit $\{a_0\}_\mathrm{max} \sim 3.0 \times 10^8 (100/\Gamma)^2 \sqrt{W_\mathrm{iso} / 10^{40}\,\mathrm{erg}}$. Synchrotron maser radiation models can also generate ultra-relativistic FRBs when the emission originates from localized plasma blobs of small volume $V$ \cite{RN98}.}

The quiver motion of an ion with mass number $A$ and charge number $Z$ is highly relativistic in an electromagnetic wave with  $a_0 \approx \sigma_i = {Am_p} / Z{m_e}$, where $m_p$ is the proton rest mass. The electromagnetic waves with  $a_0 \geq m_p/m_e \sim 1000$ are called ultra-relativistic. The interaction of an FRB pulse with charged particles is well in the ultra-relativistic regime before the burst propagates to a radius 
\begin{equation}
	R_c \approx 5.1\times{10}^{5} \left( \frac{W_\mathrm{iso}}{10^{40}\,\mathrm{erg}} \right) ^{1/2}\, \mathrm{km}. \label{equ:Rc}
\end{equation}
In the following, we investigate the particle acceleration by FRB pulses within the region $R < R_c$. At larger radii with $a_0 \gtrsim 1$ the acceleration in electron-wakefield wave driven by FRBs becomes inefficient \cite{RN26}.

{The potential progenitors of FRBs range from neutron stars and black holes to more exotic objects \cite{RN118,RN3,RN59,RN5,RN97}, yet their local environments remain poorly understood. For instance, in the magnetosphere, which may contain a background magnetic field and electron-positron pairs, ions accelerated by FRB pulses could be supplied through two channels: one extracted from the stellar surface via the central star's rotation-induced electric fields \cite{RN92, RN93, RN90} or through energetic events like flares \cite{RN108, RN94, RN98}, and another injected via accretion from the ambient medium or its binary companion \cite{RN104, RN99, RN100, RN101, RN102, RN103, RN95}.}

\section{The quasi-static and quasi-periodic plasma wake-waves}
\begin{figure}[b] 
	\centering
	\includegraphics[scale=0.85]{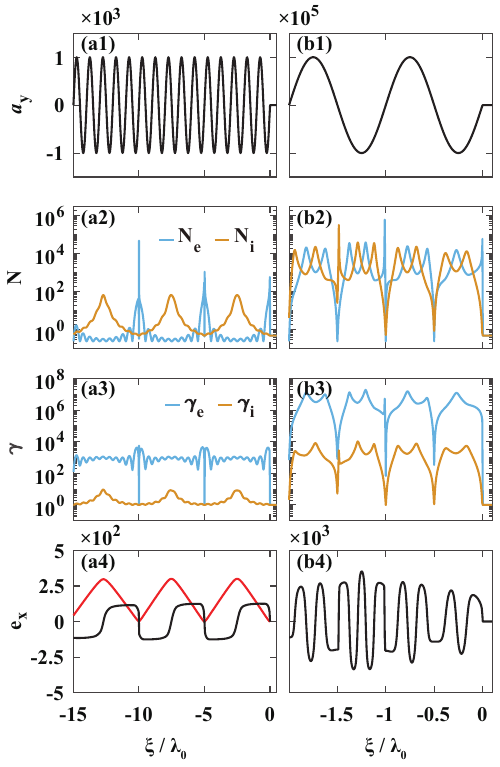}
	\caption{Two regimes of plasma wake-waves in electron--proton plasma driven by ultra-relativistic electromagnetic pulse. (a1)--(a4) are obtained for $a_0={10}^3$, (b1)--(b4) $a_0={10}^5$. (a1) and (b1) are the vector potential of the pulse. (a2) and (b2) are the normalized densities of electron and proton fluids, (a3) and (b3) are their Lorentz factors. (a4) and (b4) are the normalized longitudinal electrostatic fields as  $e_x = eE_x/m_ec\omega$. The red line in (a4) represents the scalar potential $\phi$ in arbitrary units. The plasma density for both cases is $N_0=0.5$.
		\label{fig:fig1}}
\end{figure}
We start by considering the wakefield excited by an FRB pulse in electron--ion plasma. The transverse spatial scale of an FRB pulse greatly exceeds both its own wavelength and the plasma wavelength. This permits a one-dimensional (1D) description over a certain propagation distance, where the normalized vector potential is given by $\bm{a}(x,t) = e\bm{A}(x,t)/m_ec^2$ and depends solely on the longitudinal coordinate $x$ and time $t$. In such condition, the responses of electrons and ions to an FRB pulse obey the 1D relativistic cold--fluid equations. Introducing a frame co-moving with the pulse   $\xi = x-c\beta_ct$, $\tau = t$,  with $\beta_c = v_c/c$ the co-moving velocity \cite{RN32}, and taking the quasi-static approximation ($\partial/\partial\tau=0$) \cite{RN33}, one obtains the normalized  scalar potential equation in the limit $\gamma_c^2 = 1/(1-\beta_c^2) \gg 1$ as
\begin{equation}
	k_p^{-2}\frac{d^2\phi}{d\xi^2} = \frac{1+a^2}{2\left(1+\phi\right)^2} - \frac{1+\left(a/\sigma_i\right)^2}{2\left(1-\phi/\sigma_i\right)^2}, \label{equ:equ8}
\end{equation}
where  $\phi(\xi) = e\Phi(\xi)/m_ec^2$ and $k_p = 2\pi/\lambda_p = \left(4\pi e^2n_cN_0 / m_ec^2\right) ^{1/2}$ is the plasma wavenumber, and $N_0 = n_0/n_c$ is the normalized ambient electron density with $n_c = m_e\omega^2 / 4\pi e^2 \approx 1.2\times10^{10} \mathrm{cm}^{-3}$ the critical density for FRB at 1 GHz. Once the scale potential is found from Eq.$\,$(\ref{equ:equ8}), other fluid quantities can be obtained. The normalized densities of electron and ion fluids are
\begin{align}
	\frac{N_e}{N_0} &= \frac{1+a^2}{2\left(1+\phi\right)^2}+\frac{1}{2}, \label{equ:Ne} \\ 
	\frac{ZN_i}{N_0} &= \frac{1+\left(a/\sigma_i\right)^2}{2\left(1-\phi/\sigma_i\right)^2}+\frac{1}{2}, \label{equ:Ni}
\end{align}
and the Lorentz factors are
\begin{align}
	\gamma_e &= \frac{1+a^2}{2\left(1+\phi\right)}+\frac{1+\phi}{2}, \label{equ:gamma_e} \\ 
	\gamma_i &= \frac{1+\left(a/\sigma_i\right)^2}{2\left(1-\phi/\sigma_i\right)}+\frac{1-\phi/\sigma_i}{2}. \label{equ:gamma_i}
\end{align}
When the value of $a$ is a constant with $a\left(\xi\right) \equiv a_0$ and for $\sigma_i/a_0^2 \ll 1$, an analytical solution of Eq.$\,$(\ref{equ:equ8}) can be found, which approximates a triangular wave with amplitude $\phi_{max} \approx \sigma_i$ and period $\Lambda_p \approx \sigma_i\lambda_p / \pi a_0$. {Section 1 in the Supplemental Material \cite{SM} provides a detailed derivation of the wakefield equations.}

As most FRBs are linearly polarized, we can express their vector potential as $a_y\left(\xi\right) = a_0 \sin\left(k\xi\right)$ for $-cT \le \xi \le 0$ with $k$ the wavenumber, where $cT \approx 10^6\lambda_0$. Equation$\,$(\ref{equ:equ8}) can be obtained by numerically with the Runge-Kutta method based on the Dormand-Prince (4, 5) pair \cite{RN34}. A typical solution shows in Figs.$\,$\ref{fig:fig1}(a1)--\ref{fig:fig1}(a4) for $a_0={10}^3$. The electrons with lower inertia are pushed by the intense longitudinal Lorentz force $-e{\beta_{ey}}{B}$ of the pulse, piling up to form energetic electron peaks as shown in Fig.$\,$\ref{fig:fig1}(a2). In the meanwhile, the background ions are pulled forward by the strong Coulomb force of these electron sheets, which also form energetic ion sheets. There exists quasi-periodic electron--ion sheets (Fig.$\,$\ref{fig:fig1}(a2)) and the electrostatic field (Fig.$\,$\ref{fig:fig1}(a4)) between them, which co-moves with the pulse as a wake-wave. The leading electron sheet at the pulse front ($\xi \sim 0$) is named as the front electron sheet (FES) in the following. Particles at density peaks possess extremely high Lorentz factors (Fig.$\,$\ref{fig:fig1}(a3)). Here, the ion motion is dominated by the space-charge force, placing them in the wakefield regime. %
When the FRB pulse has a higher amplitude, both electron and ion motions are governed by the Lorentz force of the pulse with a period of $\lambda_0/2$, within the piston regime. Figures$\,$\ref{fig:fig1}(b1)--(b4) are the results with $a_0=10^5$. Electrons and ions simultaneously accumulate under the Lorentz force and are shaped by the electrostatic field of the plasma wave to form a modulated density distributions, as shown in Fig.$\,$\ref{fig:fig1}(b2). The plasma wave intensifies as the peak densities, Lorentz factors, and electrostatic field strength increase, while its period reduces to less than $\lambda_0/2$ (Fig.$\,$\ref{fig:fig1}(b4)).

\begin{figure}[b]  
	\centering
	\includegraphics[scale=0.85]{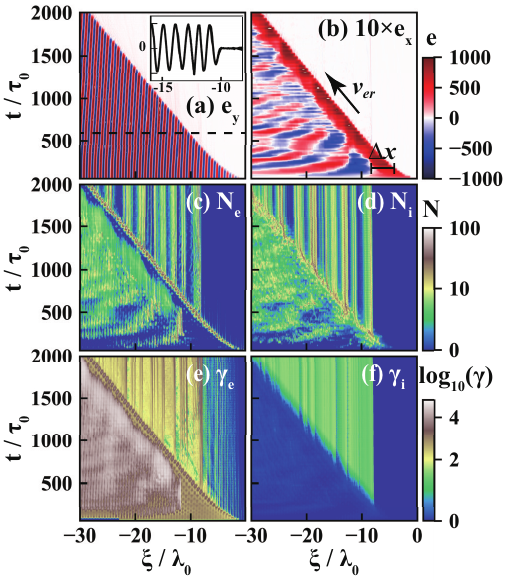}
	\caption{The acceleration process in the wakefield regime obtained from 1D-PIC simulation with $a_0={10}^3$, $\Delta\xi_\mathrm{up}=7.4\lambda_0$ and $N_0=0.5$ in electron--proton plasma. (a) Spatiotemporal evolutions of the FRB pulse front, where the inset shows the field structure at the front at $t=580 \tau_0$. (b) The normalized electrostatic field ($e_x = eE_x/m_ec\omega$) which is multiplied by a factor of 10. (c) and (d) are the electron and proton densities, respectively. (e) and (f) are their Lorentz factors. The energy spectrum of the accelerated protons shows in Fig.$\,$\ref{fig:fig4}(a).
		\label{fig:fig2}}
\end{figure}

\section{Numerical simulations of ion acceleration and energy scaling laws  in two regimes}
The above solutions only give the quasi-static structures based upon the fluid equations, which exclude the effects of the pulse energy loss, the plasma wave-breaking, and particle acceleration. These effects can be found from particle-in-cell (PIC) simulation, for example, with the code EPOCH \cite{RN35}. The simulated FRB pulse has a normalized electric field
$e_y\left(\xi\right) = a_0\tanh\left(-\xi/\Delta\xi_\mathrm{up}\right)\sin\left(k\xi\right)$ for $\xi=x-ct\le0$, where $\Delta\xi_\mathrm{up}$ represents the length of the FRB rising edge. {The simulation employed a cold, neutral electron--positron--proton plasma, with protons constituting a fraction $\mu_i \in [0, 1]$ and positrons $1-\mu_i$. The plasma density was selected within ranges consistent with potential source environments \cite{RN85,RN94,RN98}. The influences of background magnetic fields, multi-dimensional effects and radiation reaction are investigated. Detailed simulation instructions are provided in Section 2 in \cite{SM}}.

\begin{figure}[b]
	\centering
	\includegraphics[scale=0.85]{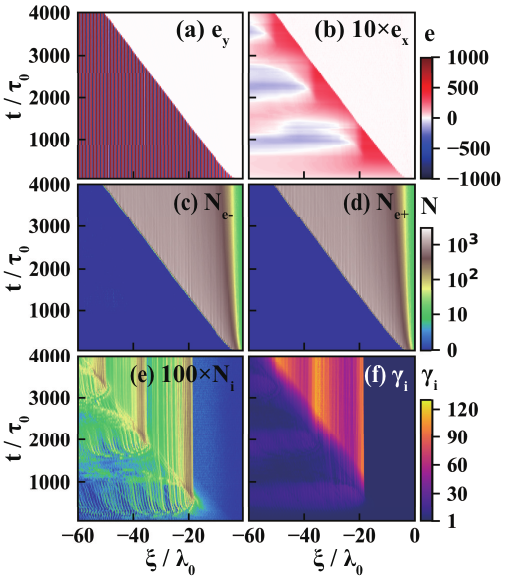}
	\caption{The acceleration process in the wakefield regime from 1D-PIC simulation with $a_0={10}^3$, $N_0=10$, $\mu_i=10^{-3}$, and $\Delta\xi_\mathrm{up}=7.4\lambda_0$ in electron--positron--proton plasma. The electrostatic field and the ion number density is multiplied by a factor of 10 and 100, respectively. The energy spectrum of accelerated protons gives in Fig.$\,$\ref{fig:fig4}(a).
		\label{fig:sm_eep}}
\end{figure}

Figure$\,$\ref{fig:fig2} illustrates the temporal evolution of both the FRB pulse front and the energetic sheets in electron--proton plasma for $a_0={10}^3$. The FRB pulse front undergoes continuous erosion by the FES, leading to the formation of a sharp front \cite{RN84}. This is evident in Figs.$\,$\ref{fig:fig2}(a)--\ref{fig:fig2}(c), where the pulse front moves backward within the moving window with the erosion velocity $v_{er}$, while the FES remains at the first half-cycle of the pulse. The proton sheet located behind the FES is continuously accelerated by the Coulomb force from the FES (Fig.$\,$\ref{fig:fig2}(b)) until this proton sheet catches up with the FES. The speed of the proton sheet is higher than the propagation velocity of the pulse, so that it moves ahead of the pulse and form a stable energetic plasma sheet with charge neutrality together with the FES. Meanwhile, the new FES is formed and the acceleration process we just discussed happens all over again. Such a process continues with the formation of multiple energetic plasma sheets, as shown in Figs.$\,$\ref{fig:fig2}(c)--(f). {In the electron--positron--proton plasma, the accumulated electrons and positrons in the FES cannot reach complete charge neutrality, i.e., the net excess of the electron density over the positron density within the FES still generates high electrostatic fields which can accelerate ions, as shown in Fig.$\,$\ref{fig:sm_eep}.}

{The kinetic energy of ions within the plasma sheets is $\mathcal{E}_\mathrm{wf} = ZecE_{x}\Delta{x}/v_{er}$, where $\Delta{x}$ is the initial distance between FES and subsequent ion sheet, $v_{er}$ is the erosion--speed of the FRB pulse front, as shown in Fig.$\,$\ref{fig:fig2}(b). These ions are accelerated by the electrostatic field $E_x$ to a velocity $\sim c$ over a duration of $\Delta{x}/v_{er}$. The wakefield equations give $\Delta x / \lambda_0 = 94\left( {a_0}\sqrt{N_0} \right)^{-0.56}$ and 
\begin{align}
	\mathcal{E}_\mathrm{wf} \approx 
	\begin{cases}
	3.0\times10^8\left(a_0/N_0\right)^{0.67} \mathrm{eV}, &\mu_i \sim 1, \\
	3.5\times10^8a_0N_0^{-0.5} \mathrm{eV}, &\mu_i \ll 1
	\end{cases}
	\label{equ:equ9}
\end{align}
for the wakefield regime, refer to Section 3 in \cite{SM} for detailed derivation.}

\begin{figure}[b]
	\centering
	\includegraphics[scale=0.85]{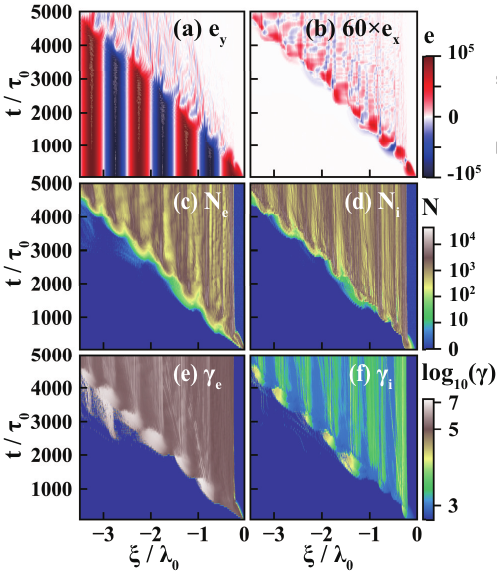}
	\caption{The acceleration process in the piston regime from 1D-PIC simulation with $a_0={10}^5$, $N_0=1$ and $\Delta\xi_\mathrm{up}=0$ in electron--proton plasma. The electrostatic field is multiplied by a factor of 60. The energy spectrum of accelerated protons gives in Fig.$\,$\ref{fig:fig4}(b).
		\label{fig:sm_pis}}
\end{figure}

As the amplitude of the FRB pulse $a_0$ increases to $10^5$, as shown in Fig.$\,$\ref{fig:sm_pis}, the immense FRB field can push all the particles along its path to accumulate within the first half-cycle of the pulse, resulting in intense pulse front dynamics. Far behind the pulse front, almost no particle left and thus no electrostatic field is formed, as shown in Figs.$\,$\ref{fig:sm_pis}(b)--(d). The particle acceleration is now in the piston regime. Compared with the wakefield regime, both the electrons and protons are accelerated simultaneously by the Lorentz force of the pulse  at the leading edge of the pulse with $\Delta{x} / \lambda_0 < 1/2$. The acceleration is governed by momentum conservation between the FRB pulse and plasma \cite{RN36,RN37,RN38,RN39}. For ultra-relativistic FRBs with $a_0^2/N_0 \gg \sigma_i$, the kinetic energy of ions in the plasma sheet is
\begin{align}
	\mathcal{E}_\mathrm{pis} &\approx 
	6.6\times{10}^8\frac{Aa_0}{\sqrt{N_0\left(2+\mu_i\sigma_i\right)}} \mathrm{eV}, \label{equ:equ12_1} \\  
	&\approx
	\begin{cases}
		1.5\times{10}^7 A^{0.5} Z^{0.5} a_0 N_0^{-0.5} \mathrm{eV}, &\mu_i \sim 1, \\
		4.7\times{10}^8 A a_0 N_0^{-0.5}\mathrm{eV}, &\mu_i \ll 1/\sigma_i.  
	\end{cases}
	 \label{equ:equ12}
\end{align}
The radiation reaction should be considered when $a_0 > 3.70\times{10}^7N_0^{-1/6}$, as the radiation reaction force approaches the Lorentz force \cite{RN46}. The acceleration process by the FRB pulse in electron-positron plasma also operates in the piston regime. See Section 4 in \cite{SM} for the detailed derivation.

\begin{figure}[b]  
	\centering
	\includegraphics[scale=0.85]{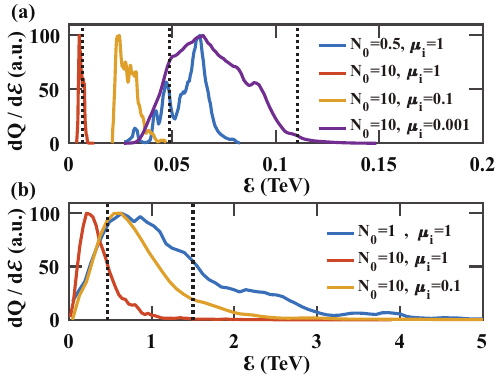}
	\caption{{The energy spectra of the accelerated protons from 1D-PIC simulations.} (a) For $a_0 = 10^3$, where the black dotted lines at 6.6$\,$GeV, 48.8$\,$GeV and 110.7$\,$GeV are the predicted values of Eq.$\,$(\ref{equ:equ9}) with $N_0=10$ \& $\mu_i=1$, $N_0=0.5$ \& $\mu_i=1$, and $N_0=10$ \& $\mu_i \ll 1$, respectively. (b) For $a_0 = 10^5$, where the black dotted lines at 0.47$\,$TeV and 1.5$\,$TeV are the predicted values of Eq.$\,$(\ref{equ:equ12_1}) with $N_0=10$ \& $\mu_i=1$, $N_0=1$ \& $\mu_i=1$ (or $N_0=10$ \& $\mu_i=0.1$), respectively.
		\label{fig:fig4}}
\end{figure}
{Figure$\,$\ref{fig:fig4} shows several proton energy spectra obtained from simulations. Each spectrum features a monoenergetic peak, matching the predictions of Eq.$\,$(\ref{equ:equ9}) and Eq.$\,$(\ref{equ:equ12_1}). Additional simulations given in \cite{SM} demonstrate that: (i) The ambient magnetic field $B_\mathrm{bg}$ near the source can confine particles to certain extent, the ultra-relativistic FRB pulse can overcome this confinement \cite{RN30} to accelerate ions to ultrahigh energies, provided its field strength significantly exceeds the background magnetic field ($E_0 \ge 10^{2\sim 3} B_\mathrm{bg}$); (ii) The acceleration process is robust against the instabilities such as filamentation according to our 2D and 3D simulations. For electromagnetic pulse with $a_0 \gg 1$, the relativistic filamentation instability is significantly suppressed \cite{RN106}.}

\begin{figure}[ht]
	\centering
	\includegraphics[scale=0.85]{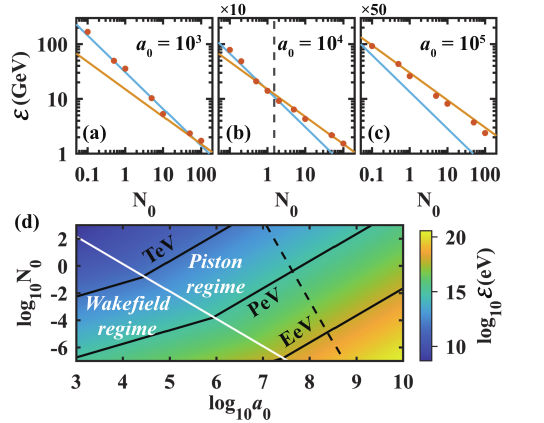}
	\caption{The kinetic energy of the protons in the accelerated plasma sheets according to theory models and 1D-PIC simulations in electron--proton plasmas. In (a)--(c), the blue lines come from Eq.$\,$(\ref{equ:equ9}), the yellow lines come from Eq.$\,$(\ref{equ:equ12}), the red dots represent the 1D-PIC simulation results. (d) is the proton kinetic energy scaling from acceleration models. The dashed line in (b) and white line in (d) represents the boundary with $\Delta{x}/\lambda_0=0.5$, which approximately separates the wakefield regime and the piston regime. Radiation reaction need to be consider at the right side of the dashed line in (d).
		\label{fig:fig5}}
\end{figure}

As shown in Figs.$\,$\ref{fig:fig5}(a)--(c), these two acceleration models describe the PIC simulation results very well. When $a_0=10^3$, the separation between neighboring sheets $\Delta{x}$ is sufficiently large, placing them in the wakefield regime. At $a_0=10^4$, an increase in $N_0$ results in a transition from the wakefield regime to the piston regime, with the boundary at $N_0 \approx 1.5$. For $a_0=10^5$, acceleration is found broadly the piston regime. Figure$\,$\ref{fig:fig5}(d) shows the proton energies from Eqs.$\,$(\ref{equ:equ9}) and (\ref{equ:equ12_1}). {The reduction of ion components ($\mu_i$) enables them to accelerate to higher energies.} Ultra-high-intensity bursts can accelerate particles to energies exceeding EeV in tenuous plasma, reaching the energy upper limit of observed CRs.

As an FRB pulse propagates outwards, its field strength decreases as $a_0 \sim 10^{8}\left(R/1\,\mathrm{km}\right)^{-1}$ from Eq.$\,$(\ref{equ:equ1}). Therefore, particle acceleration occurs in the piston regime within the radius {$R_t \sim 8.7\times10^3N_0^{0.5}$ km (where $\Delta{x}/\lambda_0 \approx 94\left( {a_0}\sqrt{N_0} \right)^{-0.56}\le 1/2$)}, and transitions to the wakefield regime in the range $R_t < R < R_c$ (where $R_c\sim 10^5\,\mathrm{km}$ from Eq.$\,$(\ref{equ:Rc})), {as shown in Fig.$\,$\ref{fig:fig0}}. The number of plasma sheets obtained through acceleration is $\mathcal{N} = \int_{1}^{R_t} dR/ct_\mathrm{pis} + \int_{R_t}^{R_c} dR/ct_\mathrm{wf}$, where $t_\mathrm{pis} = {m_i\gamma_\mathrm{pis}c}/{eE_y}$ and $t_\mathrm{wf} = \Delta{x}/v_\mathrm{er}$. Substantial particle acceleration requires $\mathcal{N} \ge 1$, which sets a lower limit for the plasma density at $7.8\times10^{-5}n_c$. On the other hand, the FES continuously erodes the burst, with the total erosion length given by $l_\mathrm{er} = \int_{1}^{R_t}(1-\beta_{FRB})dR + \int_{R_t}^{R_c}v_\mathrm{er}dR/c$. The FRB pulse is severely eroded when $l_\mathrm{er} \sim cT$, making it difficult for the FRB pulse and their accelerated particles to escape the source simultaneously. This sets an upper limit for the plasma density at {$0.1n_c$}.

The energy of the accelerated protons decreases as $a_0$ decreases, $\mathcal{E}_\mathrm{pis} \propto a_0 \propto R^{-1}$, $\mathcal{E}_\mathrm{wf} \propto a_0^{0.67} \propto R^{-0.67}$. The number of accelerated particles {$d\mathbb{N} \propto R^2dR$}. As a result, the entire expansion process naturally yields a power-law energy spectrum for the two regimes, giving by {$d\mathbb{N}/d\mathcal{E} \propto \mathcal{E}_\mathrm{pis}^{-4}$ and $\mathcal{E}_\mathrm{wf}^{-5.48}$}, respectively. The indexes close to the CR spectrum \cite{RN13}. {Considering that the background plasma density $n_0$ likely decreases with increasing distance $R$, the indexes should decrease accordingly.} The piston regime typically produces higher-energy particles than the wakefield regime, but its energy spectrum is flatter. This may lead to an ankle in the spectrum.

The total energy of accelerated particles per FRB pulse can be estimated with 
$\mathcal{E} = \int_{1}^{R_t}4\pi{n_0}R^2\mathcal{E}_\mathrm{pis}dR + \int_{R_t}^{R_c}4\pi{n_0}R^2\mathcal{E}_\mathrm{wf}dR$, which includes the contributions from the two regions for piston acceleration and wakefield acceleration. It is noted that the inferred local plasma densities surrounding FRB progenitors may span many orders of magnitude for different FRB models \cite{RN108, RN94}. In an electron--proton plasma with a density of 1 $\mathrm{cm}^{-3}$, for example, the acceleration is predominantly in the wakefield regime, as $R_t \sim 8.7\times10^3N_0^{0.5}\,\mathrm{km} \approx 79\,\mathrm{m} \ll R_c$, where $R_c\sim 10^5\,\mathrm{km}$. Equation$\,$(\ref{equ:equ9}) gives $\mathcal{E}_\mathrm{wf} \approx 3.9\times10^{20}\left(R/1\,\mathrm{km}\right)^{-0.67}\,$eV for $a_0 \sim 10^{8}\left(R/1\,\mathrm{km}\right)^{-1}$, then the total energy carried by protons accelerated beyond $10^{17}$ eV reaches about $10^{37}$ erg, about one-thousandth of the FRB pulse energy. Moreover, potential FRB sources are often embedded in intricate magnetic environments such as supernova remnants \cite{RN121}. As these relativistic particles accelerated by FRBs propagate outward, they could serve as injectors for the well-known Fermi acceleration mechanism \cite{RN119}, enabling further acceleration to even higher energy. 

\section{Conclusion}
FRBs as ultra-relativistic electromagnetic waves near their sources, can efficiently accelerate particles to ultra-high energies, making them potential directly observable messenger-type CR-accelerators. It is found that the particle energy increases with the field amplitude $a_0$ and decreases with the plasma density $N_0$. Two distinct ion acceleration regimes have been identified, i.e., the wakefield and piston regimes, each exhibiting different energy scalings. These acceleration mechanisms are also beneficial for understanding the interactions between ultra-relativistic electromagnetic pulses and plasmas in the universe \cite{RN24} and future exawatt laser facilities \cite{RN41}. The energies of accelerated particles can cover the entire cosmic ray spectrum ($10^9 \sim 10^{20}$ eV) and a power-law distribution with an index close to the CR spectrum naturally emerges as the burst expands outward. Additionally, the particle energy show a power-law dependence on both the mass number and the charge number. This acceleration is more robust than DSA \cite{RN13} in two aspects: particles can be accelerated directly by the FRB pulse from rest to relativistic speeds, i.e., no injection problem. Once accelerated, the plasma sheets propagate ahead of the burst and no longer influence subsequent acceleration. Detecting these possible high-energy counterparts of FRBs enables deeper insights into their origins and promotes the development of multi-messenger astronomy.

\begin{acknowledgments}
\textit{Acknowledgments}---We thank Dr. Huaihang Song for discussions in QED-PIC simulation, Dr. Yuanhong Qu and Dr. Mengqi Yang for discussions in FRBs, and Dr. Xiantao Cheng for discussions in the fluid equations. This work was supported by the National Natural Science Foundation of China (Grants No. 12135009, No. 12225505, and No. 12375236) and Fundamental and Interdisciplinary Disciplines Breakthrough Plan of the Ministry of Education of China (JYB2025XDXM204), and the Science Challenge Project (No. TZ2025012).
\end{acknowledgments}

\nocite{RN115,RN114,RN116,RN117,RN105,RN112,RN110,RN113,RN83,RN40}

\end{document}